\documentclass[12pt]{iopart}
\usepackage{iopams}
\usepackage{graphicx}

\begin{document}

\title[strong contamination of graphene in vacuum
systems]{Identification of a strong contamination source for
graphene in vacuum systems}

\author{Christophe Caillier$^{1,2}$, Dong-Keun Ki$^{1,2}$, Yuliya
Lisunova$^1$, Iaroslav Gaponenko$^1$, Patrycja Paruch$^1$ and
Alberto~F. Morpurgo$^{1,2}$}
\address{$^1$MaNEP-DPMC, Universit{\'e} de Gen{\`e}ve, 24 quai
Ernest-Ansermet, CH-1211 Geneva, Switzerland} \address{$^2$GAP,
Universit{\'e} de Gen{\`e}ve, 24 quai Ernest Ansermet,CH-1211
Geneva, Switzerland} \ead{\mailto{Patrycja.Paruch@unige.ch},
\mailto{Alberto.Morpurgo@unige.ch}}

\begin{abstract}
To minimize parasitic doping effects caused by uncontrolled material
adsorption, graphene is often investigated under vacuum. Here we
report an entirely unexpected phenomenon occurring in vacuum
systems, namely strong n-doping of graphene due to chemical species
generated by common ion high-vacuum gauges. The effect --~reversible
upon exposing graphene to air~-- is significant, as doping rates can
largely exceed $10^{12}$~cm$^{-2}$/hour, depending on pressure and
the relative position of the gauge and the graphene device. It is
important to be aware of the phenomenon, as its basic manifestation
can be mistakenly interpreted as vacuum-induced desorption of
p-dopants.
\end{abstract}

\maketitle

\section{Introduction}
The unique structural and electronic properties of graphene
\cite{R1,R2} have great potential for the development of
opto-electronic applications \cite{R3,R4,R5}. For instance, the
superior mechanical strength and electrical conductivity of graphene
are ideally suited for flexible electronic devices, such as foldable
tablets. Also, the gapless graphene band structure and the high
tunability of the Fermi level by gating can be exploited to realize
innovative, switchable photonic devices with a wide spectral range
(from THz to visible). The recent success in producing large area
graphene --~for instance by epitaxial growth on SiC substrate
\cite{R6} or by chemical vapor deposition (CVD)
\cite{R7,R8,R9,R10,R11,R12,R13}~-- brings all these applications
significantly closer to reality.

Further technological progress will require significant
improvements, both at the fundamental level and in very practical
aspects of the production and characterization of graphene, to
control the material, its uniformity over large areas, and its
stability over time \cite{R4,R5}. Indeed, all these aspects are
critical because the true two-dimensionality of graphene strongly
amplifies interactions with the environment, be it the supporting
substrate, or unintentionally present adsorbates. While it seems
clear that an appropriate encapsulation strategy will have to be
developed for future applications, most investigations of graphene
are currently performed under high-vacuum conditions specifically to
minimize this type of effects. In this case, desorption of
adsorbates from the graphene device is the only process expected to
occur. The frequent observation that graphene exposed to air is
strongly p-doped, whereas after insertion in a vacuum system the
doping level is significantly reduced, is often taken as evidence
for exactly such vacuum-induced desorption of adsorbates.

In this paper we report an unexpected phenomenon that strongly
influences the properties of graphene even under high vacuum
conditions, causing pronounced time instabilities, and that is
difficult to identify as it can be mistakenly attributed to
desorption of adsorbates originally present on the material.
Specifically, we find that even in high-vacuum systems, filaments of
common ion pressure gauges produce and emit active chemical species
that are adsorbed onto graphene, and that cause very large shifts in
the gate voltage $V_G$ at which the graphene charge neutrality point
occurs. We have observed the effect to be present --~and to manifest
itself very similarly~-- on graphene obtained by both CVD
\cite{R7,R8,R9,R10,R11,R12,R13} and mechanical exfoliation
\cite{R14}, with changes in density of charge carriers much larger
than $10^{12}$~cm$^{-2}$/hour. This gauge-induced doping can
overcome the original p-doping of the material as measured in air
(even when the initial doping level is very large), eventually
making the material n-doped. It can be completely reversed upon
(re)exposure of graphene to air, with no apparent damage to the
graphene or deterioration of its properties.

The phenomenon appears as long as the chemical species generated by
the ion gauge can reach the sample before being pumped away (and
therefore it sensitively depends on the position of the ion gauge in
the vacuum system relative to the graphene sample). Since vacuum
systems containing ion gauges are widely used by many research
groups working on the characterization and production of graphene,
it is important to be aware of the existence of this phenomenon, to
avoid misinterpreting results of experiments performed under vacuum
conditions. Recognizing this effect is equally important for the
further technological development of graphene synthesis and device
fabrication techniques, leading to large scale production of
graphene materials with controllable and reproducible properties.

\section{Sample preparation and measurement}
\begin{figure}
\includegraphics[width=10 cm]{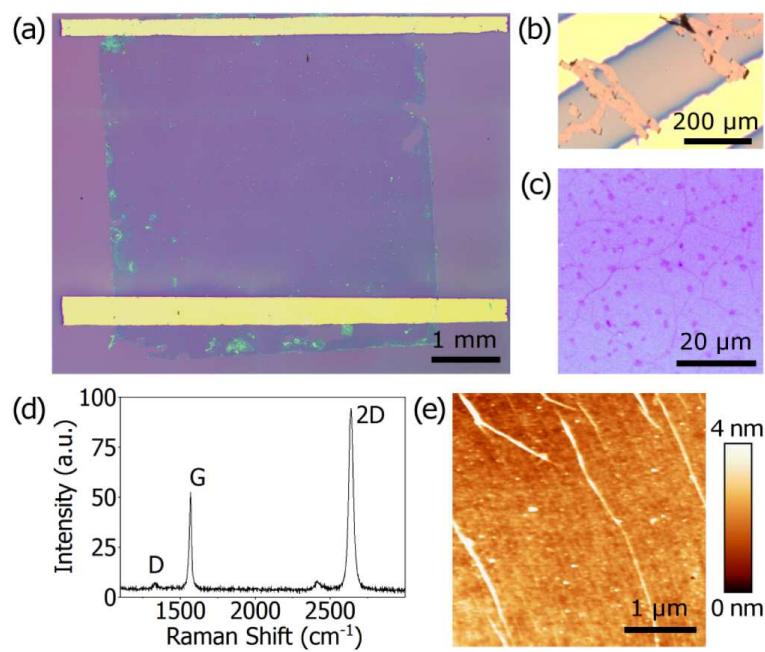}\centering
\caption[0]{(a) Large area CVD graphene sample contacted by Au/Ti
electrodes evaporated through a shadow mask, with few PMMA residues
from the transfer process visible as bright features near the edges.
(b) Smaller device fabricated by cutting graphene with a needle,
using a micro-manipulator in a probe-station. (c) High-magnification
optical microscope image of a CVD graphene sample, showing wrinkles
(dark lines) and small multilayer islands (dark spots). (d) Raman
spectrum demonstrating the monolayer character of our CVD-grown
samples. (e) Atomic force microscope of a sample transferred onto a
Si/SiO$_2$ substrate, with a roughness of $\sim$6~\AA. Small
wrinkles a few nm high (white lines) and nanoparticle residues from
the transfer process (white spots) are also visible.}
\end{figure}

Most of the graphene samples investigated here were grown by
low-pressure CVD \cite{R7,R8,R9,R10} on a 25~$\mu$m thick 99.999~\%
copper foil, following a two-step process derived from Ref.
\cite{R8}. The Cu foil is first annealed at 1000~$^{\circ}$C under a
5~sccm H$_2$ flow at 50~mTorr for 20~minutes. Graphene is then grown
in 5~sccm~H$_2$ and 7~sccm~CH$_4$ at 1035~$^{\circ}$C, with an
initial 4~minute growth step at 120~mTorr, and a second step
increasing the growth pressure to 1~Torr over 15~seconds. Finally,
the pressure is decreased back to 120~mTorr, and the samples cooled
under process gas flow for 30~minutes. These parameters have been
optimized to produce homogeneous graphene monolayers, with only few
multilayer islands or contaminating nano-particles
\cite{nanoparticles}.

Using a PMMA-mediated technique derived from Liang~\textit{et~al}.
\cite{R15}, the graphene layers were transferred onto doped silicon
substrates covered by 300~nm oxide. Under high magnification optical
microscopy (Fig. 1(c)), the samples appear uniform, with wrinkles
and small multilayer islands distinguishable, respectively, as dark
lines and spots distributed all over the sample surface.
Characterization of the sample topography by atomic force microscopy
(Fig. 1(e)) shows low $\sim$1~nm surface roughness, with small
wrinkles and nanoparticles from the transfer process. The monolayer
character of the samples is confirmed through Raman spectroscopy
measurements, shown in Fig. 1(d). A high intensity ratio between the
2D (2700~cm$^{-1}$) and G (1580~cm$^{-1}$) bands is found, as well
as a single Lorentzian profile of the 2D peak, with a width at half
maximum of 34~cm$^{-1}$ \cite{R16,R11}. The small intensity of the D
band (1350~cm$^{-1}$) demonstrates that the CVD graphene is not
disordered at small length scales \cite{R17}.

Two-terminal graphene devices were fabricated by the deposition of
source and drain electrodes (10~nm Ti / 50~nm Au) through a shadow
mask, allowing either the full samples with dimensions of about
5$\times$5~mm$^2$, or smaller 200$\times$200~$\mu$m$^2$ areas, to be
measured rapidly, without the need of any lithographic processing
(Figs. 1(a)~and~1(b); the doped Si substrate was used as gate
contact). The samples were characterized in a \textit{Lakeshore CPX}
probe station under vacuum, with a cold cathode ion gauge mounted
directly on the chamber. The resistance was obtained by
current-biasing (100~nA at $\sim$20~Hz) the graphene devices, and
measuring the source-drain voltage with an \textit{SR830} lock-in
amplifier. Over the large areas of our devices, the as-grown
material reproducibly shows a field effect mobility $\mu$ of
approximately 1000~cm$^2$V$^{-1}$s$^{-1}$ at room temperature,
increasing (also reproducibly) to approximately
1400~cm$^2$V$^{-1}$s$^{-1}$ after current annealing in vacuum. For
comparison, we also investigated an exfoliated graphene sample with
a dimension of approximately 10$\times$10~$\mu$m$^2$ (with $\mu
\sim$1800~cm$^2$V$^{-1}$s$^{-1}$, which was selected to have a
mobility value close to that of CVD-grown graphene).

\section{Results and discussion}
\begin{figure}
\includegraphics[width=10 cm]{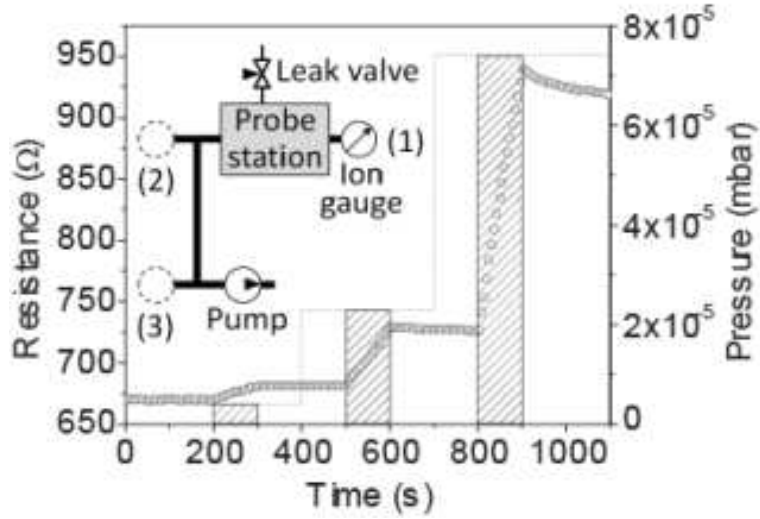}\centering
\caption[0]{Evolution of the resistance of a CVD device (open
circles) upon turning on and off the ion gauge for different values
of the pressure (dashed line) inside the chamber. The resistance
increases when the ion gauge is on (shaded regions, the height of
which corresponds to the pressure during the measurement), with a
rate found to be higher at higher pressure (see Fig. 4 and
discussion in the main text for more details). When the gauge is
switched off, the resistance remains approximately constant. Inset:
scheme of the vacuum chamber showing the position of the gauge (1)
for this measurement. Dashed circles (2) and (3) indicate
alternative positions of the gauge, displaced further away from the
sample, on the pumping line (see discussion in main text).}
\end{figure}

In order to distinguish between the effects of the ion gauge and of
vacuum itself, the chamber was first pumped with the ion gauge kept
switched off all the time, and the samples characterized at the base
pressure of 1.3$\times$10$^{-6}$~mbar. The pressure was then
stabilized at different levels up to 3$\times$10$^{-4}$~mbar, by
using a leak valve between the chamber and the ambient environment
to gradually increase the pressure in the system. For each pressure
level, the ion gauge was turned on for one to a few hundred seconds
to see its effect on the sample. As shown in Fig.~2, we observe an
increase of the sample resistance with time whenever the ion gauge
is switched on (shaded regions), and an approximately constant
resistance when the gauge is switched off. Interestingly, the rate
at which the device resistance varies upon switching on the ion
gauge appears to increase with increasing pressure inside the
chamber.

To understand the origin of the observed time dependence of the
resistance, and to determine how the rate of change depends on
pressure, we measured the conductivity of our devices as a function
of gate voltage, after having kept the devices in vacuum with the
gauge turned on for different periods of time (Fig.~3). With
increasing duration of the exposure to the ion gauge, we observe a
systematic shift of the conductivity curve towards more negative
gate voltages. Except for the slight broadening of the width of the
conductivity dip at the charge neutrality point, the shape of the
curve virtually does not change. The charge carrier mobility is thus
not affected by the ion gauge effect. One may be superficially led
to conclude that the shift originates from vacuum-induced desorption
of adsorbates attached to the graphene sheet. However, this is
clearly not the case because (i) a fast shift is only observed when
the gauge is turned on (when the gauge is turned off a shift is
still visible, but it is normally orders of magnitude slower and
smaller), and, (ii) as we discuss in detail below, the effect is
slower for better vacuum in the system. All these observations can
only be explained if the vacuum gauge generates chemical species
that propagate through the vacuum system and that act as n-dopants
once they are absorbed onto the graphene sheet.

\begin{figure}
\includegraphics[width=10 cm]{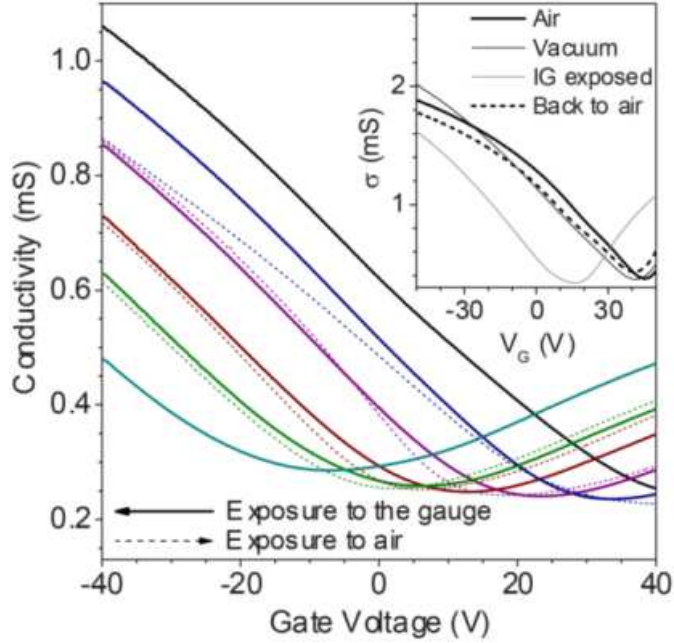}\centering
\caption[0]{Main panel: the thick solid lines -- from right (black)
to left (light blue) -- show the conductivity $\sigma$ \emph{versus}
gate voltage $V_G$, with each curve measured after having exposed
the graphene device to the ion gauge (IG) for an increasingly long
period of time. The shift of the charge neutrality point into
negative $V_G$ values indicates the occurrence of n-doping. The thin
dashed lines (from left to right) illustrate the ``recovery'' of the
sample upon exposure to air, showing the reversibility of the
effect. Inset: similar measurements performed on an exfoliated
graphene device show qualitatively similar trends (see main text).}
\end{figure}

As shown by the dashed lines in Fig.~3, the effect is reversible
after exposure to air, both in CVD graphene, previously annealed in
vacuum to improve transport characteristics (causing p-dopant
desorption, and thus a slight additional shift of the charge
neutrality point), and in exfoliated, non-annealed graphene. In the
latter, we measured the conductivity \emph{vs} gate voltage
characteristics in air, then in vacuum with the ion gauge off and
with the ion gauge on, and finally after six days of recovery in
air, as shown in the inset of Fig.~3. We see that the vast majority
of the doping effect occurs as a result of the exposure to the
gauge, and that it is completely reversed after exposure to air.

To quantify the effect of the vacuum gauge we have estimated the
doping rate $dn/dt$. The change in dopant concentration $dn$ is
obtained from the conductivity \emph{vs} gate voltage curve
($\sigma(V_G)$), using the known gate capacitance to convert the
$V_G$ shift of the charge neutrality point into a shift in carrier
density. As shown in Fig.~4 for several CVD-grown samples (open
symbols), we find that the doping rates $dn/dt$ obtained in this way
are roughly proportional to the pressure in the system, with a
coefficient of proportionality $\alpha$ of the order of
10$^{14}$~cm$^2$s$^{-1}$mbar$^{-1}$ (dashed line in Fig.~4). Devices
made with standard exfoliated graphene (filled diamonds in Fig.~4)
show qualitatively identical behavior, albeit with a slightly lower
value of the coefficient $\alpha$. We thus conclude that the
n-doping induced by the ion vacuum gauge is systematically present
for all kinds of graphene devices (i.e., not only specific to our
CVD grown samples).

The type of measurements just discussed can be used to provide a
very direct, qualitative demonstration that the effect is indeed
caused by the ion gauge. To this end, it is simply sufficient to
compare the value of the parameter $\alpha$ when repeating exactly
the same experiment, but with the vacuum gauge mounted in different
positions of the vacuum system as the only difference (the different
positions are illustrated in the inset of Fig.~2). For a specific
pressure in the vacuum system, the experimental values of $\alpha$
thus obtained are the three data points in the shaded region of
Fig.~4. The upper one (i.e. the highest $\alpha$) corresponds to the
case where the gauge is mounted directly on the probe station
(position 1 in inset of Fig.~2). $\alpha$ is an order of magnitude
lower when the gauge is mounted in front of the chamber, on the
pumping line (position 2 in inset of Fig.~2); it drops again over
two orders of magnitude when the gauge is mounted close to the
turbomolecular pump itself (position 3 in inset of Fig.~2). Thus,
although the effect is seen irrespective of the position of the
gauge in the system, its strength decreases when the gauge is
mounted further away from the sample and closer to the inlet of the
pump.

\begin{figure}
\includegraphics[width=10 cm]{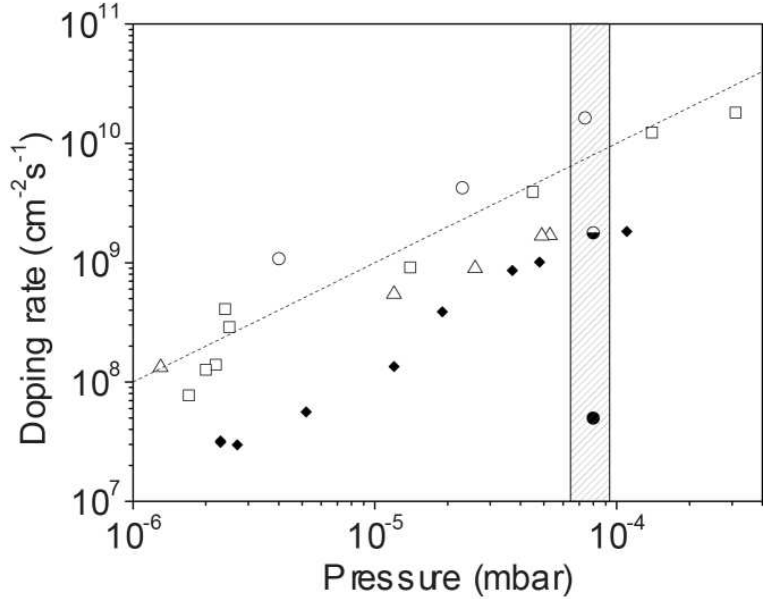}\centering
\caption[0]{Measured doping rate $dn/dt$ \emph{versus} pressure for
different samples and ion gauge positions. Open symbols denote
different CVD-grown graphene samples, and filled diamonds an
exfoliated graphene device. All the samples show pressure dependent
n-doping rates, with a coefficient of the order of
10$^{14}$~cm$^2$s$^{-1}$mbar$^{-1}$ (dashed line). The three data
points in the vertical shaded rectangle correspond to measurements
of the same CVD-grown sample at the same pressure (approximately
8$\times$10$^{-5}$~mbar), but with the ion gauge mounted at
different positions in the vacuum system, as indicated in the inset
of Fig.~2. They show how the doping rate decreases from position (1)
(open circle) to position (2) (half-filled circle) to position (3)
(filled circle). The difference is due to the efficient evacuation
of the active species created by the ion gauge when it is placed
further away from the sample and closer to the pump.}
\end{figure}

All our results are in line with observations reported earlier by
Podzorov~\emph{et~al}. \cite{R18} on a specific type of organic
single-crystal transistors, in which the surface of the crystal
where the transistor channel is formed is accessible to molecules
present in the vacuum chamber. These authors also found a reversible
and pressure-dependent n-doping generated from ion vacuum gauges,
manifesting itself in a large shift of the device threshold voltage
and --~different from the case of graphene~-- in a strong carrier
mobility suppression. Podzorov and coworkers attributed the effect
to the creation of active species, possibly free radicals, on the
hot gauge filament, which --~when adsorbed on the organic crystal~--
simultaneously dope its surface (explaining the threshold voltage
shift) and create localized electronic states in the gap of the
organic semiconductor (accounting for the suppressed mobility).

Such an explanation fully accounts for our observation on graphene
as well. Indeed, in vacuum systems, many different radicals are
known to be generated by hot filaments depending on the gas species
present \cite{R19,R20,R21,R22}. In our case, with the molecules
reaching the hot filament of the ion gauge coming from the
background pressure of our vacuum system, we certainly expect to
have radicals of O, OH, H, and of different hydrocarbons (we cannot,
however, determine if a specific radical plays a dominant role in
the observed doping of graphene, and what the microscopic mechanism
is). According to this explanation, the increase in doping rate with
increasing pressure can then be the consequence of two different
mechanisms. First, as the pressure is increased, the number of
molecules reaching the hot filament of the ion gauge is larger, and
so is the number of radicals that eventually reach the graphene
layer. Such an explanation very simply accounts for the linear
dependence between doping rate and pressure that is seen
experimentally. Additionally, at higher pressure, the mean free path
inside the vacuum system is shorter, which eventually makes the
motion of the emitted radical diffusive (rather than ballistic).
Diffusive motion should also result in an increase of the doping
rate, because the emitted radicals have a larger probability to
reach the graphene layer rather than to be adsorbed on the walls of
the vacuum system. However, we expect that such a crossover in
transport regime (from ballistic to diffusive) becomes relevant only
at pressure of the order of 10$^{-3}$ mbar or higher (out of the
range that we have investigated in detail), since at lower pressure
the molecular motion on the dimension of the vacuum systems used in
our studies is virtually always ballistic. Finally, contrary to the
case of organic semiconductors studied by Podzorov~\emph{et~al}.
\cite{R18}, graphene has no band-gap, and the only visible effect of
the chemical species generated from the ion gauge is to shift the
charge neutrality point, without significant effects on the the
carrier mobility (the mobility is likely limited by other sources of
disorder in our CVD graphene layers).

In summary, it is obviously important to be fully aware of the
effect of ion vacuum gauges on graphene, since the majority of
research groups working on this material performs experiment or
measurements in vacuum systems where these ion vacuum gauges are
very commonly employed, and kept turned on all the time. The
gauge-induced n-doping can be very misleading when it comes to
sample characterization. Graphene samples are usually p-doped if no
annealing is done prior to the measurement, and the shift of the
charge neutrality point toward smaller gate voltages could be
interpreted as p-dopant desorption in vacuum, whereas it is actually
the result of more dopants of the opposite sign produced by the ion
gauge filament, being \emph{adsorbed} onto the sample surface.
Clearly, having p-dopants desorbed or compensated by n-dopants would
lead to a very different situations if, for instance, local sample
homogeneity is a relevant issue in the experiments. Furthermore, the
vacuum gauge induced doping effect occurs most intensely at high
pressure, when the system is pumping prior to device measurement.
Once base pressure is reached, the doping rate can become small
enough to be unnoticeable over a short experimental time. It may
therefore not be obvious that the ion gauge is the actual reason why
n-doping is found in some samples.

\section{Summary}
Our observations clearly show that even in vacuum systems, where
graphene is expected to be isolated from the effect of molecules
present in the environment, adsorption of active chemicals generated
by ion pressure gauges has a significant influence on the electronic
properties of the material. Specifically, changes in carrier density
occurring at a rate of up to 10$^{13}$ cm$^{-2}$/hour at 10$^{-4}$
mbar are observed, allowing doping graphene up to a level of
$\approx 5\times10^{12}$ cm$^{-2}$. The effect is difficult to
identify (as it is easily mistakenly interpreted as vacuum-induced
desorption of adsorbates) and appears very similarly in all kinds of
graphene, both CVD-grown and exfoliated, annealed and non-annealed,
as long as the chemical species generated by the vacuum gauges can
reach the sample before being pumped out. As the doping effect takes
place in any vacuum system in which an ion gauge is used, it is
important to be aware of its existence when performing investigation
or characterization of the electronic properties of graphene.
Clearly, it will also be important to eliminate (or properly take
into account) these effects when developing a process for the
production and structuring of large-area graphene with reproducible
properties, which is a key aspect in developing a robust
graphene-based technology. Indeed, many key technological steps,
such as annealing of graphene, contact depositions, or material
etching are normally performed in vacuum systems, in which chemicals
generated by vacuum ion gauges can affect the properties of
graphene.

\section{acknowledgements}
The authors thank S. Muller and A. Ferreira for technical support,
and J. Teyssier for his support on Raman spectroscopy. This work was
funded by the Swiss National Science Foundation through the NCCR
MaNEP and Division II grant 200020-138198.


\section{References}

\begin{thebibliography}{}

\bibitem{R1} Geim A~K and Novoselov K~S 2007 {\em Nat. Mater.\/} {\bf 6} 183--191

\bibitem{R2} Castro~Neto A~H, Guinea F, Peres N~M~R, Novoselov K~S and Geim A~K
2009 {\em Rev. Mod. Phys.\/} {\bf 81} 109--162

\bibitem{R3} Geim A~K 2009 {\em Science\/} {\bf 324} 1530--1534

\bibitem{R4} Choi W, Lahiri I, Seelaboyina R and Kang Y~S 2010 {\em Crit. Rev.
Solid State\/} {\bf 35} 52--71

\bibitem{R5} Novoselov K~S, Fal'ko V~I, Colombo L, Gellert P~R, Schwab M~G and
Kim K 2012 {\em Nature\/} {\bf 490} 192--200

\bibitem{R6} Berger C, Song Z, Li T, Li X, Ogbazghi A~Y, Feng R, Dai Z,
Marchenkov A~N, Conrad E~H, First P~N and de~Heer W~A 2004 {\em The
Journal of Physical Chemistry B\/} {\bf 108} 19912--19916

\bibitem{R7} Li X, Cai W, An J, Kim S, Nah J, Yang D, Piner R, Velamakanni A,
Jung I, Tutuc E, Banerjee S~K, Colombo L and Ruoff R~S 2009 {\em
Science\/} {\bf 324} 1312--1314

\bibitem{R8} Li X, Magnuson C~W, Venugopal A, An J, Suk J~W, Han B, Borysiak M,
Cai W, Velamakanni A, Zhu Y, Fu L, Vogel E~M, Voelkl E, Colombo L
and Ruoff R~S 2010 {\em Nano Lett.\/} {\bf 10} 4328--4334

\bibitem{R9} Levendorf M~P, Ruiz-Vargas C~S, Garg S and Park J 2009 {\em Nano
Lett.\/} {\bf 9} 4479--4483

\bibitem{R10} Bae S, Kim H, Lee Y, Xu X, Park J~S, Zheng Y, Balakrishnan J, Lei T,
Ri~Kim H, Song Y~I, Kim Y~J, Kim K~S, Ozyilmaz B, Ahn J~H, Hong B~H
and Iijima S 2010 {\em Nat. Nanotechnol.\/} {\bf 5} 574--578

\bibitem{R11} Reina A, Jia X, Ho J, Nezich D, Son H, Bulovic V, Dresselhaus M~S
and Kong J 2009 {\em Nano Lett.\/} {\bf 9} 30--35

\bibitem{R12} Reina A, Thiele S, Jia X, Bhaviripudi S, Dresselhaus M, Schaefer J
and Kong J 2009 {\em Nano Res.\/} {\bf 2}(6) 509--516

\bibitem{R13} Kim K~S, Zhao Y, Jang H, Lee S~Y, Kim J~M, Kim K~S, Ahn J~, Kim P,
Choi J~ and Hong B~H 2009 {\em Nature\/} {\bf 457} 706--710

\bibitem{R14} Novoselov K~S, Geim A~K, Morozov S~V, Jiang D, Zhang Y, Dubonos S~V,
Grigorieva I~V and Firsov A~A 2004 {\em Science\/} {\bf 306}
666--669

\bibitem{nanoparticles} Small particles can be formed during graphene growth, as previously
reported in Ref. [16]. Using scanning electron microscopy, we have
additionally observed that growth conditions can modify the
nanoparticle density, size (from 100~nm to 1~$\mu$m), and shape
(from rounded to tetragonal). We suggest that, because of their
tetragonal shape, the observed nanoparticles are nanodiamonds,
although the presence of Cu residues cannot be completely excluded.

\bibitem{R15} Liang X, Sperling B~A, Calizo I, Cheng G, Hacker C~A, Zhang Q, Obeng
Y, Yan K, Peng H, Li Q, Zhu X, Yuan H, Hight~Walker A~R, Liu Z, Peng
L~m and Richter C~A 2011 {\em {ACS} Nano\/} {\bf 5} 9144--9153; Here
we used HCl 30~\% with a few drops of H$_2$O$_2$ 30~\% as the Cu
etchant.

\bibitem{R16} Ferrari A~C, Meyer J~C, Scardaci V, Casiraghi C, Lazzeri M, Mauri F,
Piscanec S, Jiang D, Novoselov K~S, Roth S and Geim A~K 2006 {\em
Phys. Rev. Lett.\/} {\bf 97} 187401

\bibitem{R17} Ferrari A~C 2007 {\em Solid State Commun.\/} {\bf 143} 47--57

\bibitem{R18} Podzorov V, Menard E, Pereversev S, Yakshinsky B, Madey T, Rogers
J~A and Gershenson M~E 2005 {\em Appl. Phys. Lett.\/} {\bf 87}
093505

\bibitem{R19} Smith J A, Cameron E, Ashfold M N R, Mankelevich Y A
and Suetin N V 2001 {\em Diamond and Related Materials} {\bf 10}
358-363

\bibitem{R20} Duan H L, Zaharias G A and Bent S F 2002 {\em
Current Opinion in Solid State and Materials Science} {\bf 6}
471-477

\bibitem{R21} Umemoto H, Ohara K, Morita D, Morimoto T,
Yamawaki M, Masuda A and Matsumura H 2003 {\em Jpn. J. Appl. Phys.}
{\bf 42} 5315

\bibitem{R22} Umemoto H and Kusanagi H 2009 {\em The Open Chem. Phys.
Journal} {\bf 2} 32-36

\end{thebibliography}
\end{document}